\documentclass[a4paper,11pt]{article}
\usepackage{pos}
\usepackage{caption}
\usepackage{subcaption}
\UseRawInputEncoding
\title{Exploring Composite Dark Matter with an SU(4) gauge theory with 1 fermion flavor}
\ShortTitle{SU(4) gauge theory with 1 fermion flavor}

\author*[a]{Venkitesh Ayyar}
\author{LSD collaboration}

\affiliation[a]{Hariri Institute of Computing and Computational science and Engineering, Boston University,\\
3 Cummington Mall, Boston, MA USA}

\emailAdd{vayyar@bu.edu}

\abstract{Several SU(N) gauge theories have been explored as candidates for producing stable dark matter particles that can explain their relative abundance, while also evading current constraints from direct, indirect and collider searches. In this talk, I will present the confinement and spectral properties of a new model we name "Hyper Stealth Dark Matter", which involves an SU(4) gauge theory with 1 quark flavor. The lightest baryon in this theory can be a potential dark matter candidate as it is protected from decay and hence can evade detection with a mass of just a few GeV. Existence of a first order confinement transition would open the possibility of potential detection of gravitational waves from such a transition at future observatories.}

\FullConference{The 40th International Symposium on Lattice Field Theory (Lattice 2023)\\
July 31st - August 4th, 2023\\
Fermi National Accelerator Laboratory\\}

\begin{document}
\maketitle

\section{Introduction}
There is increasing evidence for the existence of new matter beyond the Standard Model (SM) of particle physics termed {\it dark matter}. Evidence for this comes from several astronomical data sources such as galaxy rotation curves, cosmic microwave background, weak-lensing, etc. 
However attempts to detect dark matter directly have been unsuccessful so far. Several theoretical approaches are being pursued to explain the structure of dark matter. One such approach is {\it strongly coupled Composite Dark matter}. 

Strongly coupled Non-Abelian gauge theories are interesting in the context of Composite dark matter as they provide a mechanism to create stable particles. In QCD, the proton is stable as it is protected from decay due to the baryon number symmetry. In a similar manner, a new SU(N) gauge sector with small enough fermion content to allow confinement, will have a lightest baryon that is protected from decay. Such particles can serve as dark matter candidates. In addition, coupling of quarks in this new sector to other particles in the Standard Model will provide a mechanism for decay of composite particles in this sector.

\section{The model}
We study an $SU(4)$ gauge theory with one fermion flavor. The $ U_V(1) \times U_A(1) $ symmetry in the action is broken by the anomaly to $ U_V(1) $, which is a dark baryon number symmetry.
As a consequence, there is no chiral symmetry that can be spontaneously broken. However, given the single fermion flavor and four colors, there is a strong expectation that this theory is confining even in the absence of a chiral transition. As a consequence, one can expect composite mesons and baryons in the confined phase. 

In 2-flavor QCD, which is an SU(3) gauge theory with a symmetry breaking pattern $ SU_L(2) \times SU_R(2) \rightarrow SU_V(2) $, the pions are light as they are pseudo Nambu-Goldstone bosons. Absence of such a chiral symmetry breaking transition implies that there are no light mesons in our theory. This is interesting from the point of view of dark matter detection, as this lowers the threshold mass of dark matter candidates as there is no longer the need to raise the threshold mass of light pions to evade direct detection constraints.

It is known that first order phase transitions can generate gravitational waves ~\cite{Schwaller:2015tja,Caprini:2019egz,Caprini:2015zlo,Bailes:2021tot}. If the confinement transition in this theory is found to be first order, this would imply a potential gravitational wave signal in the early universe that can be checked with future gravitational wave detection experiments.

$ SU(4) $ gauge theories have been explored before in the context of Beyond Standard Model physics such as composite Higgs and composite Dark matter. Our collaboration studied the confinement dynamics ~\cite{LatticeStrongDynamics:2020jwi} and spectrum ~\cite{brower2023stealth} of a different $SU(4)$ gauge theory called Stealth dark matter (SDM) with 4 fermion flavors.
On the other hand, one-flavor $SU(N)$ gauge theories with $N=2$ ~\cite{Francis:2018xjd} and $N=3$ ~\cite{DellaMorte:2023ylq} have also been explored before. To the best of our knowledge, this is the first study of an $SU(4)$ gauge theory with one fermion flavor. 

Another reason for interest in this theory is that it emerges as the low energy limit of a very interesting four-flavor SU(4) gauge theory termed {\it Hyper-stealth Dark matter} (HSDM) ~\cite{hsdm:2023}. In this HSDM model, the four flavors of dark quarks are coupled to the electroweak sector of the Standard model via Yukawa couplings. This coupling to the SM implies that the composite dark particles that are not protected by any symmetries can decay to SM particles and thus provides a way for future confirmation of this theory. 
Within certain regions in the space of the couplings in this HSDM theory, scale separation of dark quark masses can be achieved to get one light dark quark and three heavy quarks. Moreover, the lightest quark turns out to be electrically neutral thereby causing the lightest baryon to also be electrically neutral as well as have zero charge radius, magnetic moment and polarization. This further assists in evading direct detection constraints. This is analogous to QCD if the neutron were lighter than the proton.

\section{Simulation details}

For simulating this theory, we use Mobius-Domain wall fermions. Although the extra dimension adds to the computational costs, this formulation avoids explicitly breaking the $ U_A(1) $ symmetry. 
We use the Grid software framework\footnote{https://github.com/paboyle/Grid}~\cite{Boyle:2016lbp} for generating gauge configurations and the Hadrons framework \footnote{https://github.com/aportelli/Hadrons}~\cite{antonin_portelli_2023_8023716} for measurements.
For the gauge sector, we use the plain Wilson gauge action and the Exact one-flavor algorithm~\cite{201455} for gauge generation.
For this initial exploratory study, we show simulation results on lattice sizes $ 16^3 \times 8 $ and $ 24^3 \times 12$.

As a first step, we explore the thermodynamics of the theory. The two free parameters in this theory are the gauge coupling $\beta$ and the bare quark mass $ m_f $. For our initial exploration, we fix $ m_f = 0.1 $

In this exploratory study, we compute two observables: the { \it Plaquette} and the { \it Polyakov loop}. For a pure gauge theory, the polyakov loop is an order parameter for the breaking of the centre symmetry and hence the confinement transition. For this theory however, 
since fermions explicitly break the centre symmetry, the polyakov loop is a pseudo-order parameter. However, one expects it to show discontinuity near the transition.

For these runs we choose the size of the 5th Domain-wall dimension to be 16. 
This choice is to ensure that the residual mass is close to $ 10^{-5} $. We use periodic boundary conditions in both space and time.

In this one-flavor theory, the lightest particles are the meson of type $ \bar{q} q $ and baryon $ q \ q \ q\ q$. Computing the mass of this meson is quite challenging as it also gets contributions from disconnected diagrams similar to the $ \eta $ meson in QCD.

The gauge field consists of an $ SU(4) $ matrix on each link connecting adjacent sites on the lattice. 
We perform simulations of two types with two different initial gauge configurations: {\it hot start} with random gauge links and { \it cold start} with unit gauge links. These corresponds to the two extreme cases: { \it hot start} corresponds to the limit $ \beta \rightarrow 0 $ and { \it cold start} corresponds to the limit $ \beta \rightarrow \infty $. It is expected that results from both these starts will be identical within uncertainties. In the region near the transition however, one expects some deviation with some hysteresis.
\section{Results}

\clearpage
\begin{figure}[htb!]
\includegraphics[width=0.8\linewidth]{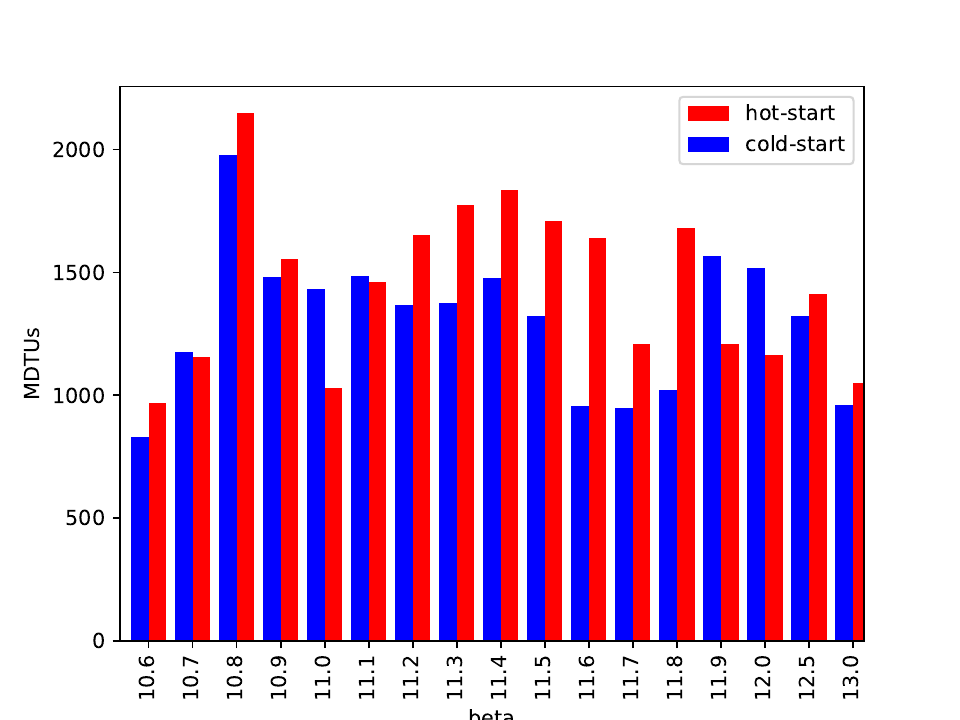}
\caption{The number of MDTUs(molecular dynamics time units) generated for each value of $ \beta $. }
\label{fig:num_conf}
\end{figure}

\begin{figure}[htb!]
\includegraphics[width=0.8\linewidth]{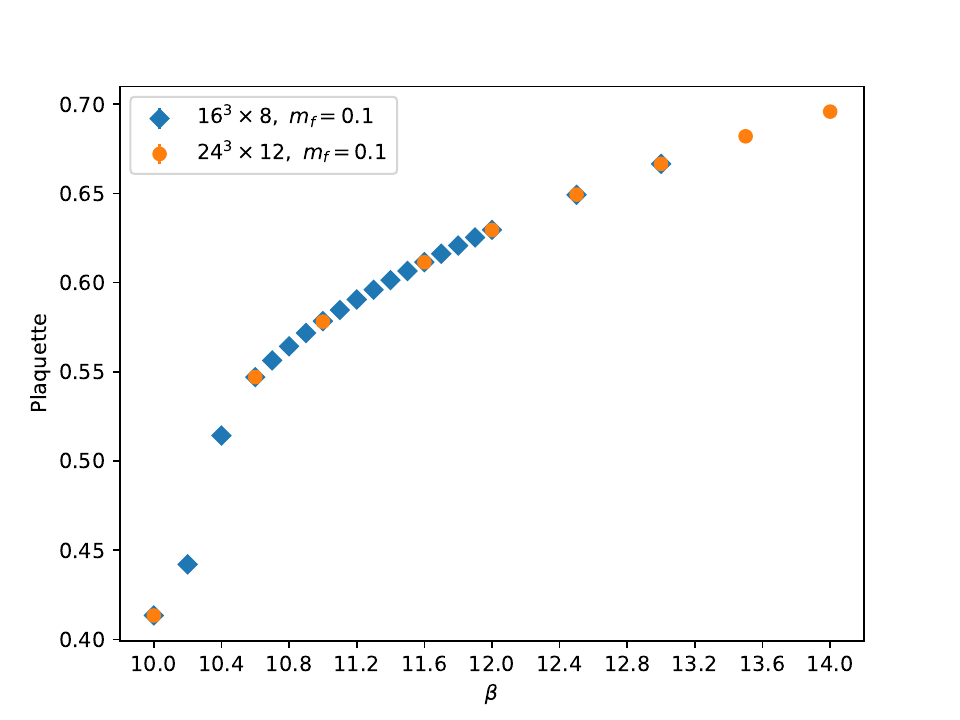}
\caption{ Variation of Plaquette with $ \beta $ for two aspect ratios $ 16 ^ 3 \times 8 $ and $24^3 \times 12$. }
\label{fig:plaq}
\end{figure}

\clearpage
\begin{figure}[htb!]
\includegraphics[width=.8\linewidth]{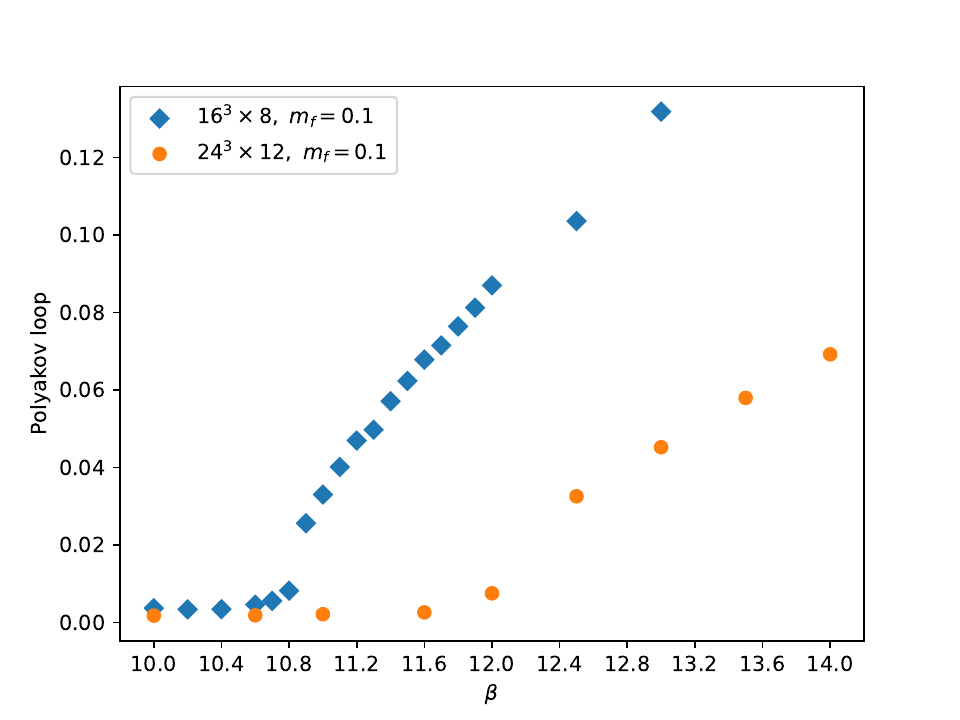}
\caption{ Variation of the Polyakov loop with $ \beta $ for two aspect ratios $ 16 ^ 3 \times 8 $ and $24^3 \times 12$. The discontinuity in the Polyakov loop is seen in the region $ \beta \sim 10.9 $ for $ 16^3 \times 8$ lattices and $ \beta \sim 12.2 $ for $ 24^3 \times 12 $.}
\label{fig:poly}
\end{figure}

\begin{figure}[htb!]
\includegraphics[width=.8\linewidth]{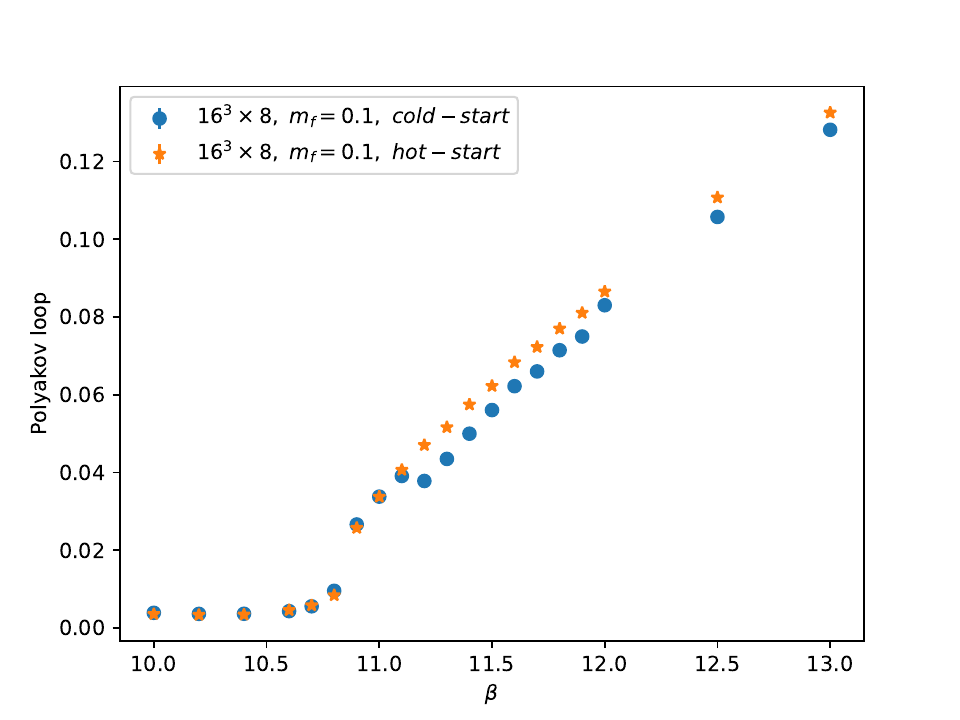}
\caption{ Variation of the Polyakov loop with $ \beta $ for $ 16 ^ 3 \times 8 $ lattices for both hot and cold starts. The values agree up until $ \beta = 11.1 $. Deviation between the two values shows that we do not have enough configurations at these values of $ \beta $.}
\label{fig:poly_hot_cold}
\end{figure}

To locate the confinement phase transition, we look at the behavior of the { \it Plaquette } and the { \it Polyakov loop}. While the Plaquette is not an order parameter, it is very useful to locate lattice artifacts such as bulk transitions, which are of no physical interest. On the other hand, a jump in the Polyakov loop can signal the existence of a confinement phase transition. 

In Fig. ~\ref{fig:num_conf} we show the number of gauge configurations generated in units of molecular dynamics time units (MDTUs) for various values of the coupling $ \beta $ for hot and cold start runs. It can be seen that the number of MDTUs obtained in this exploratory analysis is quite small, at around 1000-2000. (The study in ~\cite{LatticeStrongDynamics:2020jwi} used upwards of 10,000 MDTUs for points near the transition.)

First, we look at the behavior of the Plaquette as a function of $ \beta $ in Fig. ~\ref{fig:plaq} for the two lattice sizes $ 16^3 \times 8$ and $ 24^3 \times 12$. It can been seen that plaquette shows a discontinuous jump around $ \beta = 10.4$. This value doesn't change as we increase the temporal extend $N_t$. Hence we can conclude that this discontinuity is a lattice artifact and commonly known as a bulk transition.

In Fig. ~\ref{fig:poly} we see the variation of the polyakov loop as a function of $ \beta $,  for the two lattice sizes $ 16^3 \times 8$ and $ 24^3 \times 12$. It can be seen that the discontinuity in the polyakov loop shifts to the right from around $ \beta \sim 10.9 $ for $ 16^3 \times 8 $ to $ 24^3 \times 12 $. This is consistent with expectations for a confinement transition, since at the critical temperature $ T = \frac{1}{a N_t} $, an increase in the temporal extend $ N_t $ corresponds to decrease in the lattice spacing $ a $, which implies a shift towards a large coupling $ \beta $.
In Fig.~\ref{fig:poly_hot_cold}, we contrast this for hot and cold starts. The values for the two starting configurations agree until $ \beta = 11.1$. Beyond that, we do see a discrepancy primarily because we do not have enough configurations for both hot and cold starts.

\begin{figure}[htb!]
\includegraphics[width=.8\linewidth]{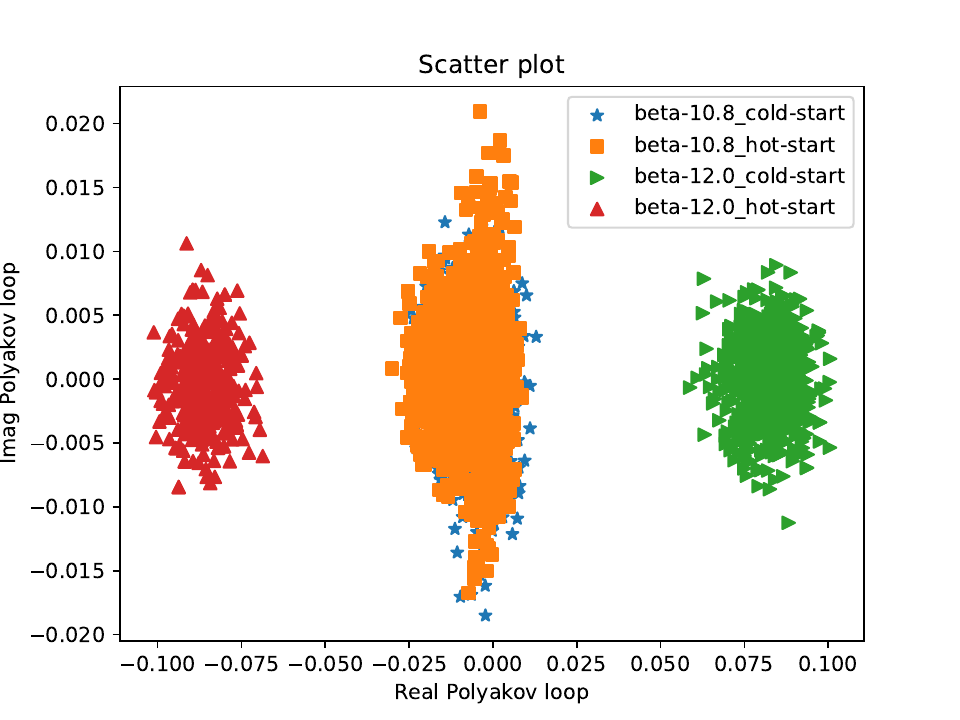}
\caption{ Scatter plot showing both the real and imaginary parts of the Polyakov loop for two values of the coupling $ \beta $, one each in the confined and deconfined phases for a lattice of size $ 16^3 \times  8 $. The Polyakov loop in the deconfined phase saturates to different non-zero values for the hot and cold starts, due to the periodic boundary conditions and limited statistics.}
\label{fig:poly_scatter}
\end{figure}

It it interesting to look at the complex value of the polyakov loop in more detail. In Fig.~\ref{fig:poly_scatter}, we show the scatter of polyakov loop in the complex plane for both hot and cold starts for two values of $ \beta $, one in the confined phase and the other in the deconfined phase.
For $ \beta=10.8$, the polyakov loop scatters around zero for both starts. For $ \beta=12.0 $, the value saturates to different non-zero values for hot and cold starts, due to the periodic boundary conditions and limited statistics.







\section{Summary and Future direction}
We have shared some early results in our exploratory study of the finite-temperature confinement transition of an SU(4) gauge theory with one fermion flavor, which is of interest as a composite dark matter model. Our exploration of thermodynamics points to the existence of a confinement transition.

Our current results show that to determine the critical coupling precisely, we need to generate many more configurations in the region close to the transition. We also need to repeat these calculations at a few different masses. 

There are plenty of avenues to explore with this theory. 
An important goal is to obtain the order of the transition by computing the Polyakov loop susceptibility and track its variation as a function of lattice size. If the transition happens to be first order, this would indicate the potential of a primordial gravitational wave signal from this confinement transition in the early universe, should this theory be realized in nature.

Our next step would be to study the spectroscopy of this theory, starting with the masses of the lightest meson and baryon.  The ratio of these masses would be a scale-independent prediction that could be of interest in phenomenology. We also intend to compute the scattering properties of this baryon.

\begin{acknowledgments}
This work used the computational resources at Lawrence Livermore National Laboratory. V.A acknowledges the support of the DOE Exascale Computing Project (ECP).

\end{acknowledgments}


\bibliography{main}{}
\bibliographystyle{ieeetr}

\end{document}